\documentclass[preprint,aps ,nofootinbib]{revtex4}
\usepackage{graphicx}
\pdfoutput=1
\usepackage{epsfig}
\usepackage{amsmath}
\usepackage{amsfonts}
\usepackage{amssymb}
\usepackage{color}
\usepackage{dcolumn}
\usepackage{hyperref}
\usepackage{multirow}
\usepackage{booktabs}
\providecommand{\U}[1]{\protect\rule{.1in}{.1in}}

\newcommand{\f}{\begin{equation}}
\newcommand{\ff}{\end{equation}}
\newcommand{\fa}{\begin{eqnarray}}
\newcommand{\ffa}{\end{eqnarray}}

\begin{document}
\title{Astrophysical observables for regular black holes with sub-Planckian curvature }
\author{Wei Zeng$^{1}$}
\email{cengwei0702@stu.cwnu.edu.cn}
\author{Yi Ling $^{2,3,1}$}
\email{lingy@ihep.ac.cn}
\author{Qing-Quan Jiang$^{1}$}
\email{qqjiangphys@yeah.net} \affiliation{$^1$ School of Physics and Astronomy, China West Normal University, Nanchong 637002, China \\$^2$ Institute of High Energy
Physics, Chinese Academy of Sciences, Beijing 100049, China\\$^3$
School of Physics, University of Chinese Academy of Sciences,
Beijing 100049, China}

\begin{abstract}
 We investigate the photon sphere and the extremal stable circular orbit (ESCO) for massive particles over a recently proposed regular black holes with sub-Planckian curvature and Minkowskian core. We derive the effective potential for geodesic orbits and determine the radius of circular photon orbits, with an analysis on the stability of these orbits. We extend our analysis to the background of compact massive object (CMO) without horizon, whose mass is below the lowest bound for the formation of a black hole. For massive particles, the ESCOs become double-valued in CMO phase and we calculate the innermost stable circular orbits (ISCO) and the outermost stable circular orbits (OSCO). By comparing wih Bardeen black hole and Hayward black hole, it is also found that the locations of photon sphere and ESCO in CMO phase with Minkowskian core are evidently different from the ones in CMO phase with dS core, which potentially provides a way to distinguish these two sorts of black holes by astronomical observation.

\end{abstract}
\maketitle
\section{Introduction}
It had been a challenging problem to identify a compact massive object to be a black hole by astronomical observation until  recently remarkable progress was made by the Event Horizon Telescope (EHT) collaboration team. The successful detection of black holes at the center of M87 and Sgr A* has announced the coming of a new age for the study of black holes as genuine astrophysical objects\cite{EventHorizonTelescope:2019dse}\cite{EventHorizonTelescope:2022xnr}. In this direction, detecting the photon sphere closely surrounding the black hole silhouette plays a key role in measuring various parameters of black holes, such as the size, the mass as well as the spin of black holes.

It is well known that the existence of black hole solutions in general relativity also brings some fundamental problems which are notoriously hard to solve in theoretical physics.  Among of them the most famous one is the singularity problem, which tells us that the scalar curvature becomes divergent at the center of black holes\cite{Hawking:1966sx,Penrose:1964wq,Joshi:2011rlc,Goswami:2005fu,Janis:1968zz}. This trouble becomes severe after the Hawking radiation was discovered, which leads to the evaporation of  black holes and finally the information loss paradox\cite{Hawking:1976ra,Giddings:1992hh,Hawking:1975vcx,Preskill:1992tc,Xiang:2013sza,Chen:2014jwq,Casadio:2000py,Unruh:1976db,Page:1993wv}. Theoretically, people believe that the infinity of the Kritchmann scalar curvature just implies the classical general relativity would not be applicable to the space time with extremal environment, and the singularity at the center of a classical black hole might be removed or avoided by the quantum effects of gravity\cite{Garay:1994en,tHooft:1984kcu,Han:2004wt,Donoghue:1993eb,Callan:1992rs,DeWitt:1967yk,DeWitt:1967ub,DeWitt:1967uc,Calmet:2017qqa,Ali:2015tva}. Before a complete theory of quantum gravity could be established, it is very desirable to construct some black hole solutions which are singularity free.  In literature, such regular black holes were constructed at the phenomenological level indeed, which now may be classified by their asymptotic behavior near the center of black holes. One sort of regular black holes has asymptotically de-Sitter core, including the famous Bardeen black hole, Hayward black hole as well as the Frolov black hole\cite{Bardeen:1968,Hayward:2005gi,Frolov:2014jva}; while the other sort of regular black holes has asymptotically Minkowskian core, which is characterized by an exponentially suppressed Newton potential\cite{Xiang:2013sza,Ling:2021olm,Li:2016yfd,Ben-Amots:2011lbe,Simpson:2019mud,Culetu:2013fsa,Martinis:2010zk}. In particular, a new class of regular black holes with asymptotically Minkowskian core was constructed recently in \cite{Ling:2021olm}, whose scalar curvature is not only finite, but also maintains to be sub-Planckian during the course of evaporation, irrespective of the mass of the black hole. Such a scenario is quite compatible with the expectation of quantum gravity, which asserts that the energy scale of any object should be bounded by the Planck energy in the context of quantum theory of gravity. In addition, the connection between regular black holes with asymptotically dS core and those with asymptotically Minkowskian core has been disclosed in \cite{Ling:2021olm}. Recently, more features of this sort of black holes have been investigated in  \cite{Ling:2021abn,Ling:2022vrv}.

Currently, although it is still far to provide a complete understanding on these regular black holes, it is very intriguing to link these regular black holes to astrophysical observation and explore if there exists any evidence or signature implying that the  detected black holes in the sky would be regular black hole rather than the traditional black hole with singularity. Obviously, at the current stage it is impractical to distinguish singular black holes and regular black holes by diving into the horizon or detecting any signal coming out from the interior of the horizon. Nevertheless, the difference inside black holes may leave signatures on the featured phenomena outside black holes, such as the photon sphere or the trajectories of massive particles\cite{Claudel:2000yi,Boonserm:2018orb,Berry:2020ntz,Decanini:2009mu,Stefanov:2010xz,Wei:2011zw,Wei:2013mda,Cunha:2018acu}. With this purpose, in this paper we intend to provide detailed analysis on the photon sphere and the extremal stable circular orbit (ESCO) for massive particles over recently proposed regular black holes with sub-Planckian curvature, and expect our theoretical investigation will provide basis and useful information for distinguishing different kinds of black holes by astrophysical observation in near future.

The paper is organized as follows. In next section we will present a short review on regular black holes with sub-Planckian curvature which are characterized by two factors $(x,n)$. We will focus on the structure of its horizon, which usually contains two branches, namely, the inner horizon and the outer horizon. When the mass is below some value, the horizon will disappear and the black hole becomes a horizonless compact massive object. In section three and four, we consider two regular black holes with  $(x=2/3, n=2)$ and $(x=1,n=3)$, respectively, and determine the radius of circular photon orbits, with an analysis on the stability of these orbits. For massive particles, the curve of ESCO becomes double-valued in CMO phase and we calculate the innermost stable circular orbits (ISCO) and the outermost stable circular orbits(OSCO). By comparing with those in Bardeen black hole and Hayward black holes, we find that the locations of photon sphere and ESCO over the horizonless CMO with Minkowskian core are evidently different from the ones over the CMO with dS core, which potentially provides a way to distinguish these two sorts of black holes by astronomical observation.

\section{The regular black hole with sub-Planckian Kretschmann scalar curvature}
In Ref.\cite{Ling:2021olm},  a new sort of regular black holes with sub-Planckian Kretschmann scalar curvature was proposed with the following metric

\begin{equation}\label{Eq.metric}
d s^{2}=-(1+2 \psi(r)) d t^{2}+\frac{1}{(1+2 \psi(r))} d r^{2}+r^{2} d
\Omega^{2},
\end{equation}
where $\psi(r)$ is understood as the modified Newton potential which is specified as
\begin{equation}\label{NP}
\psi(r)=-\frac{M}{r}e^{\frac{-\alpha_0M^x}{r^n}}.
\end{equation}
The modified Newton potential is characterized by an exponentially suppressing form with a deviation parameter $\alpha_0$. Obviously, when $\alpha=0$,  it goes back to the standard Schwarzschild black hole. The behavior of Kretschmann scalar curvature at the core depends on both  factors $x$ and $n$. In Ref.\cite{Ling:2021olm}, it has been pointed out that to guarantee the existence of the horizon and the scalar curvature to be sub-Planckian, two factors must satisfy the conditions $n\geq x \geq n/3$ and $n\geq 2$. Furthermore, as $r\rightarrow 0$, the space time is characterized by a Minkowskian core, in contrast to the well-known regular black holes such as Bardeen black hole, Hayward black hole as well as Frolov black hole, which are characterized by a de-Sitter core. Nevertheless, a one-to-one correspondence between such regular black holes with Minkowskian core and the ones with de-Sitter core has been constructed in \cite {Ling:2021olm}. Explicitly, given a regular black hole with Minkowskian core which is described by Eq.(\ref{NP}), then correspondingly there exists a regular black hole with de-Sitter core whose Newton potential $\psi(r)$ is given by
\begin{equation}
\psi(r)=-\frac{M r^{\frac{n}{x}-1}}{(r^n+x \alpha_0 M^{x})^{1/x}}.
\end{equation}
Specially, Bardeen black hole is just given by specifying $x=2/3,n=2$, while Hayward black hole is given by $x=1,n=3$. These two sorts of regular black holes exhibit identical asymptotic behavior at large scales but possess different cores at the center. Therefore, it is quite intriguing to explore practical ways to distinguish them by possible astronomical observation in future. Here as the starting point we will investigate this issue from the theoretical point of view by comparing the distinct behavior of photon spheres and ESCO of massive particles over such two sorts of backgrounds.

Now we present some basic properties of the regular black holes with the metric given by Eq.(\ref{Eq.metric}) and Eq.(\ref{NP}). In general, the location of the horizon is determined by $1+2\psi(r_h)=0$, namely
\begin{equation}
2M=r_he^{{\alpha_0M^x}/{r_h^n}}.
\end{equation}
From this equation, one may obtain  the radius of the horizon $r_h$ as the function of the mass
\begin{equation}
r_H=\left[-\frac{M^xn\alpha_0}{W(-2^{-n}M^{x-n}n\alpha_0)}\right]^{\frac{1}{n}},\label{LH}
\end{equation}
where $W$ is the Lambert-W function which allows two possible branches of solutions with appropriate values of $\alpha_0$, corresponding to the outer horizon and the inner horizon, respectively.

Next we consider the geodesics of photons and massive particles over such a background.
Firstly, since there are two Killing vectors, namely $(\partial /\partial t)^a$ and $(\partial /\partial \phi)^a$ for the space time,  correspondingly there are two conservative quantities for freely moving particles, namely the energy $E$ and the angular momentum $L$. Now it is straightforward to derive  the effective potential
for geodesic orbits as \begin{equation}
V_\epsilon(r)=\left(1+2\psi\right)\left(\frac{L^2}{r^2}-\epsilon\right),
\end{equation}
where $\epsilon=0$ for photons and $\epsilon=-1$ for massive particles. As a result, the location of the photon sphere $r_c$ is determined by the vanishing of the derivative of the effective potential with respect to the radial coordinate, giving rise to

\begin{equation}
-3Mr_c^n+e^{M^xr_c^{-n}\alpha_0}r_c^{1+n}+M^{1+x}n\alpha_0=0.\label{PS}
\end{equation}
In parallel, for massive particles the location of ESCO is denoted as $r_{e}$ which is determined by the equation
\begin{equation}\label{ES}
\begin{aligned}
    &r_{e}^{2n}\left(-6M+e^{M^{x}r_{e}^{-n}\alpha_{0}}r_{e}\right)-M^{x}n r_{e}^{n}\left[2M(-4+n)-e^{M^{x}r_{e}^{-n}\alpha_{0}}n r_{e}\right]\alpha_0 \\&-M^{2x}n^{2}\left(2M+e^{M^{x}r_{e}^{-n}\alpha_{0}}r_{e}\right)\alpha_0^2=0.
\end{aligned}
\end{equation}
The derivation of this equation can be found in next sections for special x and n.

Following the strategy presented in \cite{Berry:2020ntz}, we may define two dimensionless quantities as
\begin{equation}
\omega=\frac{r}{\alpha_0^{1/(n-x)}};\quad\quad\quad z=\frac{M}{\alpha_0^{1/(n-x)}},
\end{equation}
then Eq.(\ref{LH}), Eq.(\ref{PS}), Eq.(\ref{ES}) can be rewritten as
\begin{equation}
\omega_{H}=2z\left[\frac{\theta}{W(\theta)}\right]^{\frac{1}{n}}\quad;\quad \theta=-\frac{n}{2^n z^{n-x}},
\end{equation}
and
\begin{equation}
-3z\omega_c^n+e^{z^{x}\omega_c^{-n}}\omega_c^{1+n}+z^{1+x}n=0,
\end{equation}
and
\begin{equation}
\begin{aligned}
    &\omega_{e}^{2n}\left(-6z+e^{z^{x}\omega_{e}^{-n}}\omega_{e}\right)-z^{x}n \omega_{e}^{n}\left[2z(-4+n)-e^{z^{x}\omega_{e}^{-n}}n \omega_{e}\right] \\
    &-z^{2x}n^{2}\left(2z+e^{z^{x}\omega_{e}^{-n}}\omega_{e}\right)=0.
    \end{aligned}
\end{equation}
It means that under the condition $x\neq{n}$, we may plot the locations of horizon, the photon sphere and the ESCO as the functions of $z$, which exhibit a universal behavior independent of the deviation parameter $\alpha_0$. Nevertheless, one can easily derive the effects of the deviation parameter with the use of parameters  $M=z\alpha_0^{\frac{1}{n-x}}$ and $r=\omega\alpha_0^{\frac{1}{n-x}}$.

\section{The regular black hole with $x=2/3$ and $n=2$}
In this section we investigate the photon sphere and ESCO over the regular black hole with $x=2/3$ and $n=2$, which corresponds to Bardeen black hole at large scales. The thermodynamic behavior of this black hole as well as the feature of Kretschmann scalar curvature has been investigated in detail in \cite{Ling:2021olm}.
\begin{figure} [htbp]
  \center{
 \includegraphics[width=0.6\textwidth,height=0.6\textwidth]{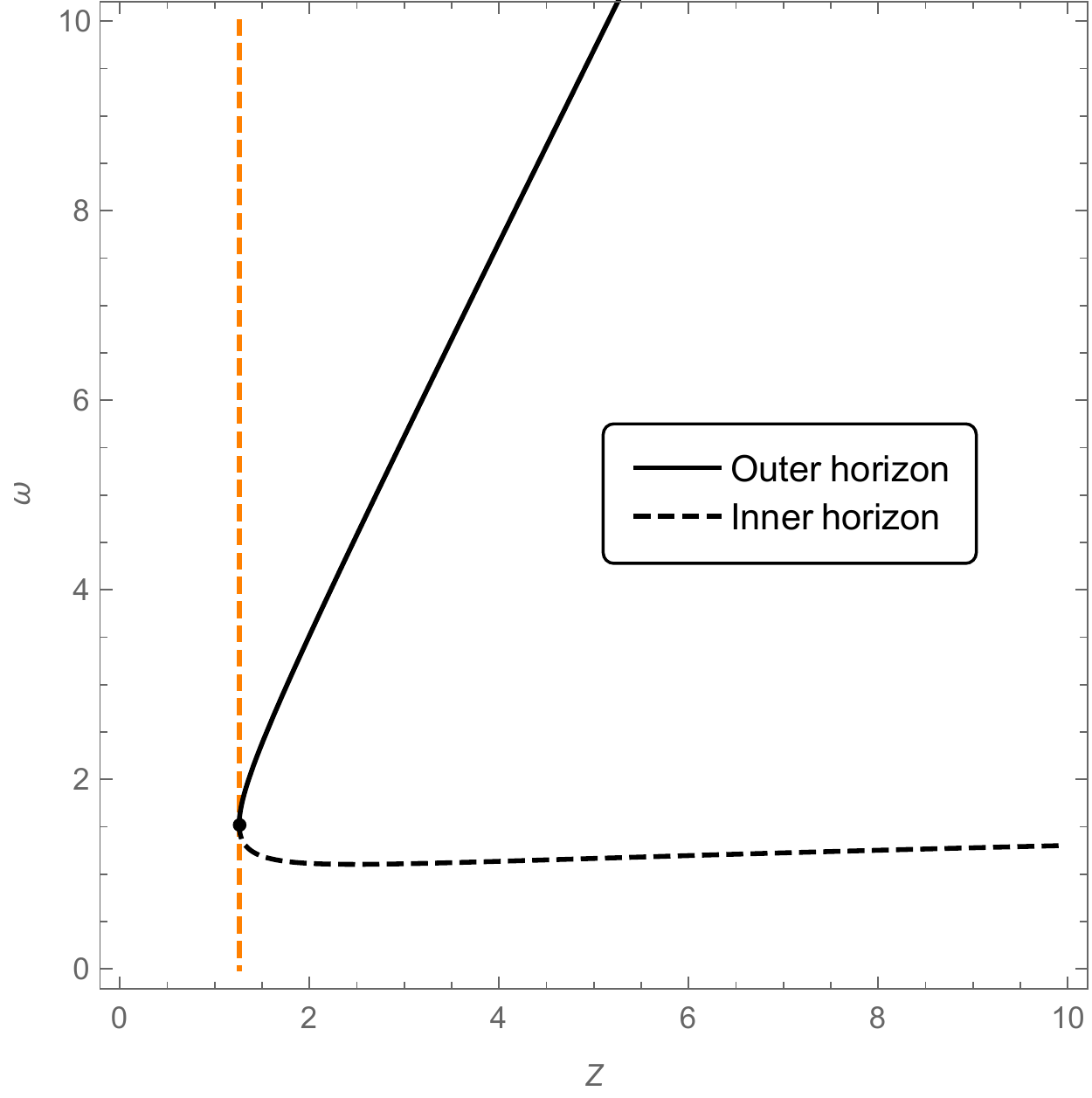} \hspace{0.05cm}
  \caption{\label{fig1} The outer horizon (black solid line) and the inner horizon (black dashed line) of the black hole with $x=2/3$ and $n=2$. The vertical dashed line in orange is the borderline between the region in black hole phase (right) and the region in horizonless CMO phase (left).} }
\end{figure}

Firstly, we present the basic properties of the regular black hole with $x=2/3$ and $n=2$. The location of the horizon is given by Eq.(\ref{LH}). Remind that Lambert-W function has two branches in the real region\cite{Corless:1996zz,Valluri:2000zz,Valluri:2009zz}, which give rise to the inner horizon and the outer horizon  separately for this regular black hole as
\begin{equation}
\omega_{H^+}=2z\left[\frac{\theta}{W_{0}(\theta)}\right]^{1/2};\quad\quad
\omega_{H^-}=2z\left[\frac{\theta}{W_{-1}(\theta)}\right]^{1/2};\quad\quad
\theta=-\frac{1}{2z^{4/3}}.
\end{equation}
We plot the inner horizon and outer horizon as the function of the mass in Fig.(\ref{fig1}). From this figure one notices that with the decrease of the mass, the inner horizon and the outer horizon become closer and finally merge at some critical point. As a matter of fact, the location of such a critical point can be evaluated by the feature of Lambert-W function as follows.

Two branches of Lambert-W function merge at $W_{-1}(\theta)=W_{0}(\theta)$, implying that $W(\theta)=-1$.
Using the identity of Lambert-W function
\begin{equation}
    \theta=W(\theta)e^{W(\theta)},
\end{equation}
one finds the inner horizon and outer horizon merge at $\theta= -1/e$, namely
\begin{equation}
 z=\frac{M}{\alpha_0^{3/4}}=\left(\frac{e}{2}\right)^{3/4}=1.258...;\quad \omega= \frac{r}{\alpha_0^{3/4}}=1.526....
 \end{equation}

As a result, the condition for the existence of the horizon is $\theta\geq -1/e$, namely
\begin{equation}
z\geq \left(\frac{e}{2}\right)^{3/4},
\end{equation}
below of which the horizon disappears such that the black hole becomes a CMO. It is interesting to notice that this situation is in contrast to the standard Schwarzschild black hole, where the horizon does not disappear as the mass decreases, but shrinks to zero with its mass.
\subsection{Photon Spheres}
\begin{figure} [htb]
  \center{
  \includegraphics[width=0.6\textwidth,height=0.6\textwidth]{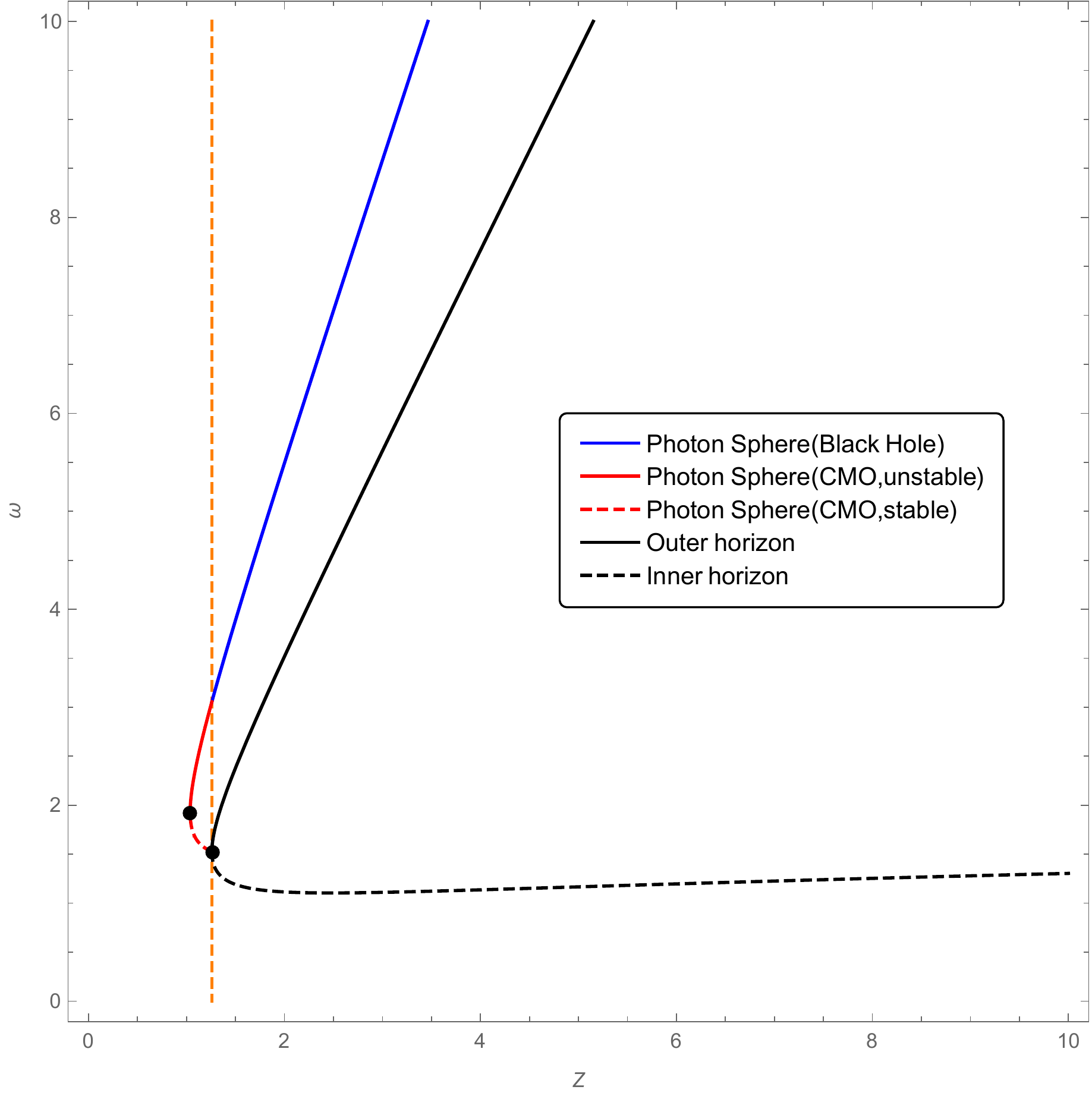}\ \hspace{0.05cm}
   \caption{\label{fig2} The location of photon spheres in the spacetime with $x=2/3$ and $n=2$. The blue line represents the location of photon spheres in black hole phase, while the red solid line is the location of unstable photon spheres in CMO phase, and the red dashed line is the location of stable photon spheres in CMO phase. The turning point of each curve is marked by a black dot as well.}}
\end{figure}

Next we consider the property of photon spheres over this background. The effective potential is given by
\begin{equation}
V_\epsilon(\omega)=\left(1-\frac{2ze^{-\frac{z^{2/3}}{\omega^2}}}{\omega}\right)\left(\frac{L^2}{\omega^2}-\epsilon\right).
\end{equation}
For null trajectories, the effective potential is
\begin{equation}
V_0(\omega)=\left(1-\frac{2ze^{-\frac{z^{2/3}}{\omega^2}}}{\omega}\right)\left(\frac{L^2}{\omega^2}\right).
\end{equation}
Thus, the location of the circular photon orbit can be determined by $V_{0}'(\omega_{c})=0$, leading to
\begin{equation}\label{Eq42}
-3z\omega_{c}^2+e^\frac{z^{2/3}}{\omega_{c}^2}\omega_{c}^3+2z^{5/3}=0.
\end{equation}
We numerically plot the radius of the photon sphere as the function of the mass in Fig.(\ref{fig2}). First of all, we notice without surprise that the radius of the photon sphere is always larger than the outer horizon.  In particular, if we rewrite the equation for the photon sphere in terms of  $M$ and $r$, then it becomes
 \begin{equation}
     -3Mr_{c}^{2}+e^{\frac{M^{2/3}\alpha_0}{r_{c}^{2}}}r_{c}^3+2M^{5/3}\alpha_0=0.
 \end{equation}
Obviously, when $\alpha_0$ is vanishing, the location of the photon sphere is $r_c=3m$, a well-known result for Schwarzschild black hole.
Secondly, we find in the horizonless CMO phase, the radius of the photon sphere becomes double-valued, as illustrated in Fig.(\ref{fig2}). This phenomenon is consistent with the results presented in \cite{Cunha:2017qtt,Guo:2020qwk}, where it is found that photon spheres always appear in pairs outside a horizonless compact object. The location of the turning point is given by $dz/d\omega_{c}=0$, namely
\begin{equation}
\frac{dz}{d{\omega_{c}}}=\frac{2e^\frac{z^{2/3}}{\omega_{c}^2}z^{2/3}+6z\omega_{c}-3e^\frac{z^{2/3}}{\omega_{c}^2}\omega_{c}^{2}}{10z+2e^\frac{z^{2/3}}{\omega_{c}^2} \omega_{c}-9z^{1/3}\omega_{c}^2}=0.
\end{equation}
Numerically one finds  the coordinates for the turning point are $(z=1.037..., \omega=1.929...)$. \par

As a summary of this subsection, we argue that the photon spheres exhibit abundant structure in this background. In the region of black hole phase $(z\geq 1.258)$, the radius of the photon sphere is single-valued and approaches $3M$ as $\alpha_0$ goes to zero. In the region of CMO phase with $(1.258 \geq z\geq 1.037)$, the radius of the photon sphere has two solutions with the same mass; while in the region with $(z < 1.037)$, we find no photon sphere exists.

\subsection{The stability of photon spheres}
Next we are concerned with the stability of photon spheres, which is related to the second order of the derivative of the effective potential. The stable photon orbit locates at a place where $V''_{0}(\omega_{c})\geq 0$,
\begin{equation}
   V''_{0}(\omega_{c})= \frac{2e^{-\frac{z^{2/3}}{\omega_{c}^2}}L^2\left(-12z\omega_{c}^{4}+3e^{\frac{z^{2/3}}{\omega_{c}^2}}+18z^{5/3}\omega_{c}^2-4z^{7/3}\right)}{\omega_{c}^{9}}\geq 0.
\end{equation}
So, we have
\begin{equation}\label{Eq43}
-12z\omega_{c}^4+3e^\frac{z^{2/3}}{\omega_{c}^2}\omega_{c}^5+18z^{5/3}\omega_{c}^2-4z^{7/3}\geq 0.
\end{equation}
One can figure out the stable region of photon orbits by numerically solving this equation. Combining Eq. ({\ref{Eq42}}) and Eq. ({\ref{Eq43}}), we can get the demarcation point of whether the photon sphere orbit is stable or unstable, as illustrated in Fig.(\ref{fig2}).
We find that the photon sphere is not stable in the region of  black hole phase. While in the region of CMO phase, the upper branch is unstable, but the lower branch is stable.

\subsection{Circular orbits for massive particles}

For massive particles, we set $\epsilon=-1$ in the effective potential
\begin{equation}
V_{-1}(\omega)=\left(1-\frac{2z e^{-\frac{z^{2/3}}{\omega^2}}}{\omega}\right)\left(\frac{L^2}{\omega^2}+1\right).
\end{equation}
So, the radius of circular orbits can be determined by $V'_{-1}(\omega_{e})=0$, leading to
\begin{equation}
    z\omega_{e}^{4}-2\omega_{e}^{5/3}z^{2}-L^{2}\left(-3z\omega_{e}^{2}+e^{\frac{z^{2/3}}{\omega_{e}^2}}\omega_{e}^3+2z^{5/3}\right)=0.
\end{equation}
With the above equation the angular momentum can be written as
\begin{equation}\label{Eq53}
L^2=\frac{z\omega_{e}^{4}-2\omega_{e}^{5/3}z^{2}}{-3z\omega_{e}^{2}+e^\frac{z^{2/3}}{\omega^{2}}\omega_{e}^{3}+2z^{5/3}}.
\end{equation}
\begin{figure} [htbp]
  \center{
 \includegraphics[width=0.6\textwidth,height=0.6\textwidth]{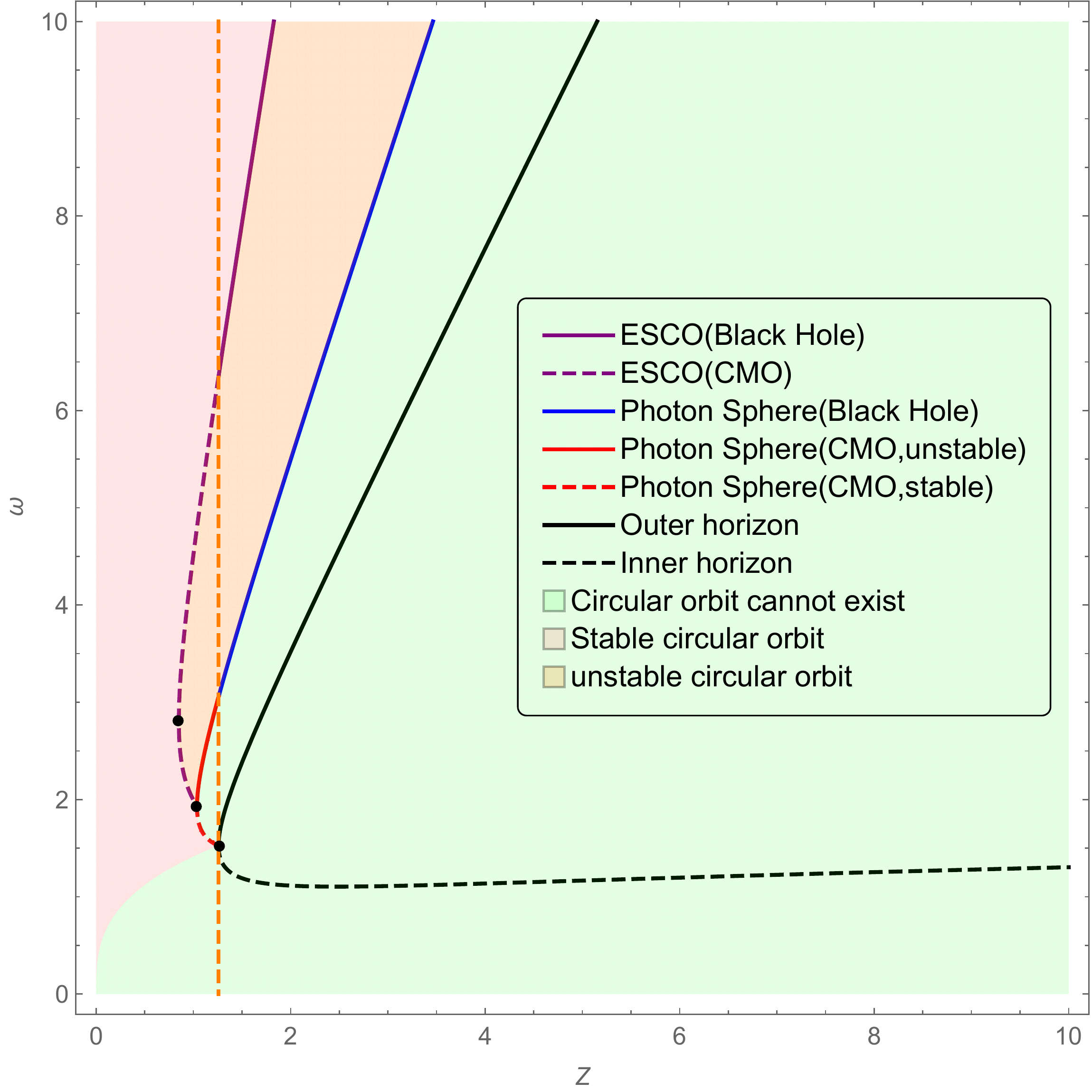}\ \hspace{0.05cm}
   \caption{\label{fig3} The location of ESCO for massive particles in the spacetime with $x=2/3$ and $n=2$. The purple solid line represents the location of ESCO in black hole phase, while the purple dashed line is the ESCO in CMO phase, where the upper branch is ISCO and the lower branch is OSCO.  In the region shaded in orange, all the circular orbits are unstable, while in the region shaded in  pink, all the circular orbit are stable. The region shaded in green denotes the place where the circular orbit does not exist. The turning point of each curve is marked by a black dot as well.}}
\end{figure}

The positivity of the angular momentum $(0 \leq L^2 < \infty)$ constrains the conditions for the existence of the circular orbit to be
\begin{equation}
    z\omega_{e}^{4}-2\omega_{e}^{5/3}z^2 \geq 0 ;\quad -3z\omega_{e}^{2}+e^{\frac{z^{2/3}}{\omega_{e}^2}}\omega_{e}^{3}+2z^{5/3} >0.
\end{equation}

Furthermore, the stability of these circular orbits is determined by the second derivatives of the effective potential, namely $V_{-1}''(\omega_{e})\geq 0$, which requires
\begin{equation}\label{Eqec}
-\omega_{e}\left(\omega_{e}^{4}+4z^{2/3}\omega_{e}^{2}-4z^{4/3}\right)+e^\frac{-z^{2/3}}{\omega_{e}^{2}}\left(6z\omega_{e}^{4}-8z^{5/3}\omega_{e}^{2}+8z^{7/3}\right) \geq 0.
\end{equation}
The equal sign in the above equation gives rise to the location of ESCO which is plotted as a curve in $(z, \omega)$ plane, as illustrated in Fig.(\ref{fig3}).
In this figure we notice that in black hole phase, the ESCO is single-valued and it is an ISCO; while in the region of CMO phase, there are two possible ESCO locations for a fixed value of $z$. In this case the upper branch is an ISCO and the lower branch is an OSCO. Moreover, the curve of ESCO does not terminate at the horizon, but at the photon sphere.

The turning point of the curve of ESCO can be obtained by solving $dz/d\omega_{e}=0$, which gives rise to
\begin{equation}
    16z^{3}-16z^{3/7}\omega_{e}^{2}+4e^{\frac{z^{2/3}}{\omega_{e}^2}} z^{4/3}\omega_{e}^{3}-4z^{5/3}\omega_{e}^{4}-12 e^{\frac{z^{2/3}}{\omega_{e}^2}}z^{2/3}\omega_{e}^{5}+24z\omega_{e}^{6}-5e^{\frac{z^{2/3}}{\omega_{e}^2}}\omega_{e}^{7}=0
\end{equation}
The numerical analysis indicates that the turning point is located at
\begin{equation}
    z=\frac{M}{\alpha_{0}^{3/4}}=0.850...;\quad \omega=\frac{r}{\alpha_{0}^{3/4}}=2.660....
\end{equation}
At the turning point, the ESCO turns from an ISCO into an OSCO.

As the summary of this subsection, we remark that the circular orbits for massive particles exist only in the region outside photon pheres. In particular, those between the curve of ESCO and the curve of photon sphere are unstable, while those outside the curve of ESCO are stable, as illustrated in Fig.(\ref{fig3}).

\begin{figure} [htbp]
  \center{
  \includegraphics[width=0.6\textwidth,height=0.6\textwidth]{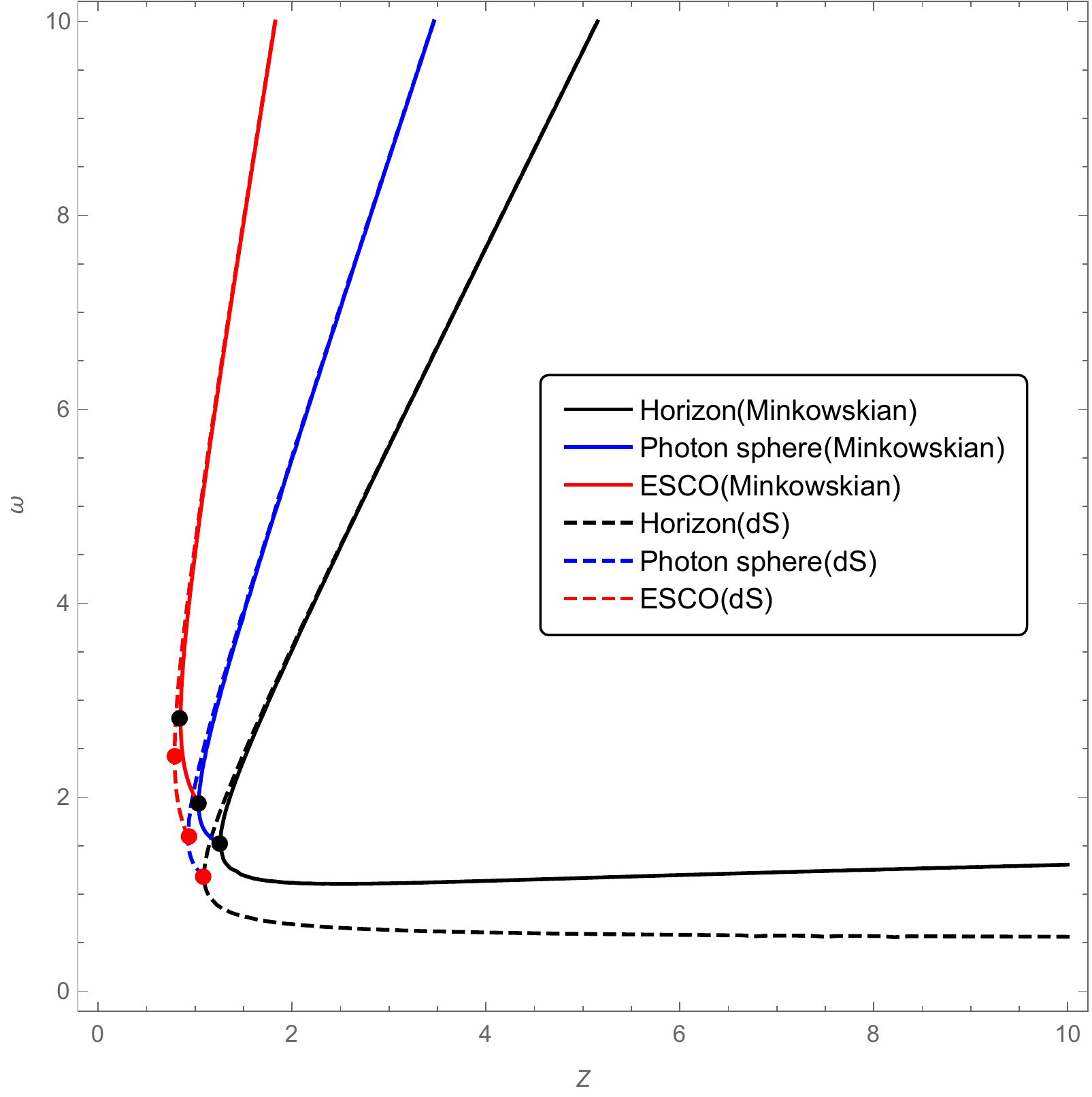}\ \hspace{0.05cm}
  \caption{\label{fig4}  The comparison of photon spheres and ESCO in the spacetime with Minkowskian core $(x=2/3, n=2)$  and those in Bardeen spacetime with dS core.}}
\end{figure}
\subsection{In comparison with Bardeen black hole}
It is quite instructive to compare the photon sphere as well as ESCO over the spacetime with $(x=2/3, n=2)$ and those over Bardeen spacetime, since these two spacetimes have the same asymptotic behavior at large scales in radial direction.

Bardeen black hole has a dS core. The previous investigation on photon orbits over Bardeen black hole can be found in \cite{Stuchlik:2014qja}. With the same value of the deviation parameter $\alpha_0$, the effective potential is given by
\begin{equation}
V_\epsilon(\omega)=\left[1-\frac{2z\omega^2}{\left(\omega^2+\frac{2}{3}z^{2/3}\right)^{3/2}}\right]\left(\frac{L^2}{\omega^2}-\epsilon\right).
\end{equation}
Thus the radius of the photon circular orbits is determined by $V'_0(\omega_{c_{2}})=0$, namely
\begin{equation}\label{Eq26}
-\frac{2}{\omega_{c_2}^3}+\frac{54\sqrt3 z\omega_{c_2}}{\left(3\omega_{c_2}^2+2z^{2/3}\right)^{5/2}}=0.
\end{equation}
For massive particles, the ESCO can be obtained in a parallel way as we demonstrate for the black hole with $x=2/3$ and $n=2$, which is given by
\begin{equation}\label{Eq27}
  \begin{aligned}
    &64 z^2 \left({2z^{2/3}+3\omega_{e_2}^2}\right)^{1/2}-162z^{2/3}\omega_{e_2}^{4}\left(2z^{2/3}+3\omega_{e_2}^{2}\right)^{1/2}\\&-27\omega_{e_2}^{6}\left[\left(2z^{2/3}+3\omega_{e_2}^{2}\right)^{1/2}-6\left(3\right)^{1/2}z\right]=0.
  \end{aligned}
\end{equation}
We compare the radius of photon spheres as well as ESCO in these two spacetimes with the same deviation parameter in Fig.(\ref{fig4}).
In black hole phase, it is hard to distinguish the regular black hole with $(x=2/3, n=2)$ and Bardeen black hole by detecting either the photon sphere or the ESCO, since these curves are almost overlapped. Nevertheless, one may theoretically analyze their differences by expanding the radius of the photon sphere in series of the deviation parameter $\alpha_0$ for these two black holes. For Bardeen black holes, one finds
\begin{equation}
    r_{c_2}=3M+\frac{5}{9M^{1/3}}\alpha_{0}+\mathcal{O}[\alpha_0]^2,
\end{equation}
while for the black hole with $(x=2/3,n=2)$, one has
\begin{equation}
    r_{c}=3M-\frac{5}{9M^{1/3}}\alpha_0+\mathcal{O}[\alpha_0]^2.
\end{equation}
Firstly, in comparison with the standard result in Schwarzschild black hole, which is $3M$, we find the radius of the photon sphere in Bardeen black hole becomes larger, while the radius of the photon sphere in black hole with $(x=2/3, n=2)$ becomes smaller. In particular, with the decrease of the mass, we find the discrepancy becomes larger. We remark that the above result coincides with the result in \cite{Ling:2022vrv}, where the shadow of these regular black holes are compared and it is found that the shadow of the black hole with $(x=2/3, n=2)$ is more deformed and the size is smaller. We may also expand the radius of ESCO in two spacetimes near $\alpha \rightarrow 0$.
For Bardeen black holes, one finds:
\begin{equation}
    r_{e_2}=6M+\frac{19}{18M^{1/3}}\alpha_0+\mathcal{O}[\alpha_0]^{2}.
\end{equation}
For the black hole with $(x=2/3, n=2)$, one has
\begin{equation}
    r_{e}=6M+\frac{19}{3M^{1/3}}\alpha_0+\mathcal{O}[\alpha_0]^{2}.
\end{equation}
We find both radii are larger than $6M$, but the radius of ESCO in the black hole with $(x=2/3, n=2)$ is a little bit larger than that in Bardeen black hole.
In the region of CMO phase, which could be viewed as the remnants of the evaporation of these two black holes, the behavior of these curves are different and in principle can be distinguished by the observation, as demonstrated in Fig.(\ref{fig4}).
From Eq.(\ref{Eq26}) and Eq.(\ref{Eq27}), one can obtain the turning points of the curves of photon sphere and ESCO by $dz/d\omega_{e}=0$ for Bardeen spacetime. We collect the locations of these turning points in two different space times in table (\ref{table1}).
\section{The regular black hole with $x=1$ and $n=3$}
In this section we investigate photon spheres and ESCO for massive particles in the spacetime with $x=1$ and $n=3$, which corresponds to Hayward black hole at large scales. Its thermodynamic behavior  as well as the feature of Kretschmann scalar curvature can be found in \cite{Ling:2021olm}.

In this background, the outer horizon and the inner horizon are separately given by
\begin{equation}
\omega_{H^+}=2z\left[\frac{\theta}{W_{0}(\theta)}\right]^{1/3};\quad \omega_{H^-}=2z\left[\frac{\theta}{W_{-1}(\theta)}\right]^{1/3};\quad \theta=-\frac{3}{8z^2}.
\end{equation}
\begin{figure} [t]
  \center{
  \includegraphics[width=0.6\textwidth,height=0.6\textwidth]{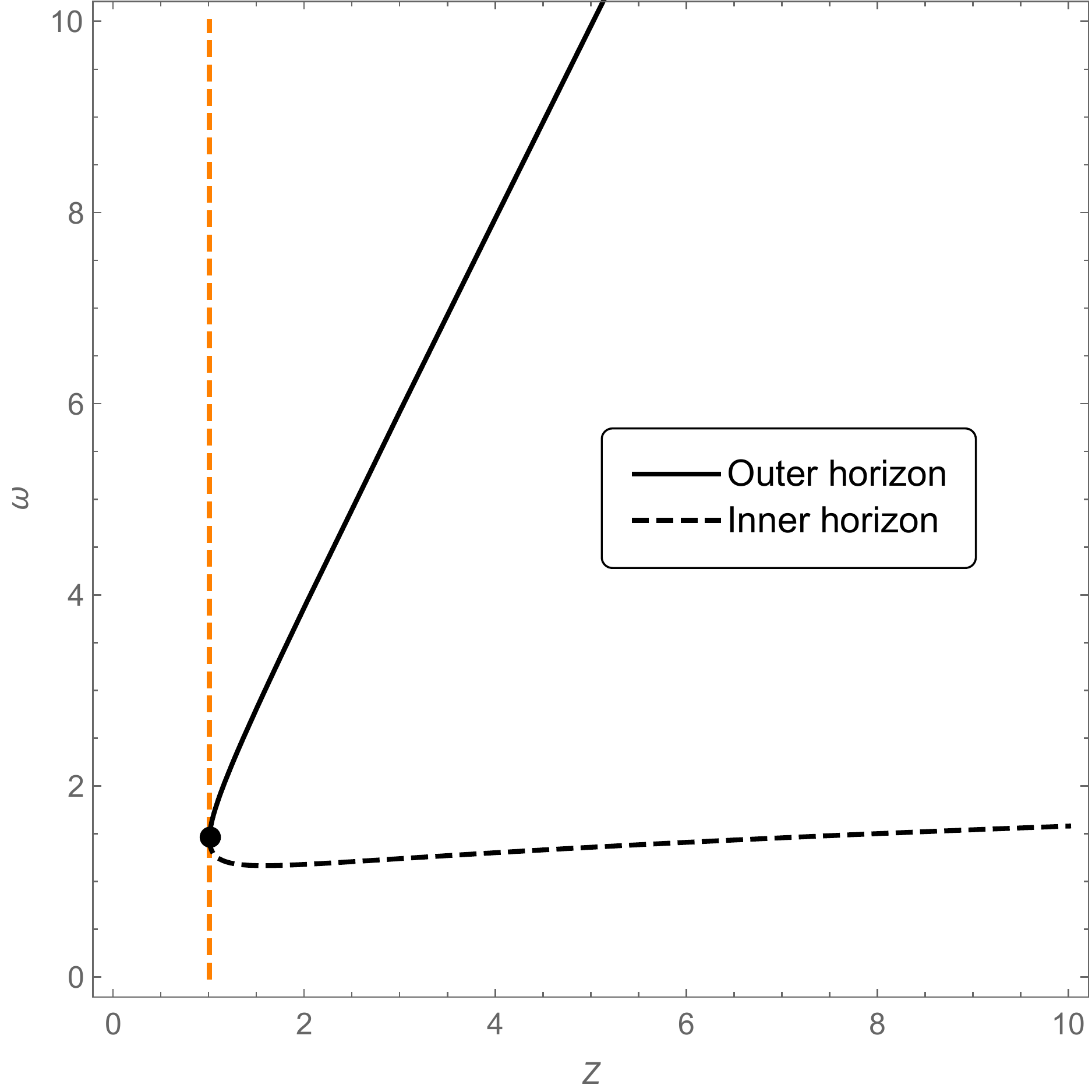}\ \hspace{0.05cm}
  \caption{\label{fig5}  The outer horizon (black solid line) and the inner horizon (black dashed line) of the black hole with $x=1$ and $n=3$. The vertical dashed line in orange is the borderline between the region in black hole phase (right) and the  region in CMO phase (left).} }
\end{figure}

We plot the locations of the outer horizon and the inner horizon in Fig.(\ref{fig5}). The point where the inner horizon and the outer horizon meet is given by
\begin{equation}
z=\frac{M}{\alpha_0^{1/2}}=1.010...;\quad \omega=\frac{r}{\alpha_0^{1/2}}=1.465....
\end{equation}

\subsection{Photon Spheres}
\begin{figure} [t]
 \center{
  \includegraphics[width=0.6\textwidth,height=0.6\textwidth]{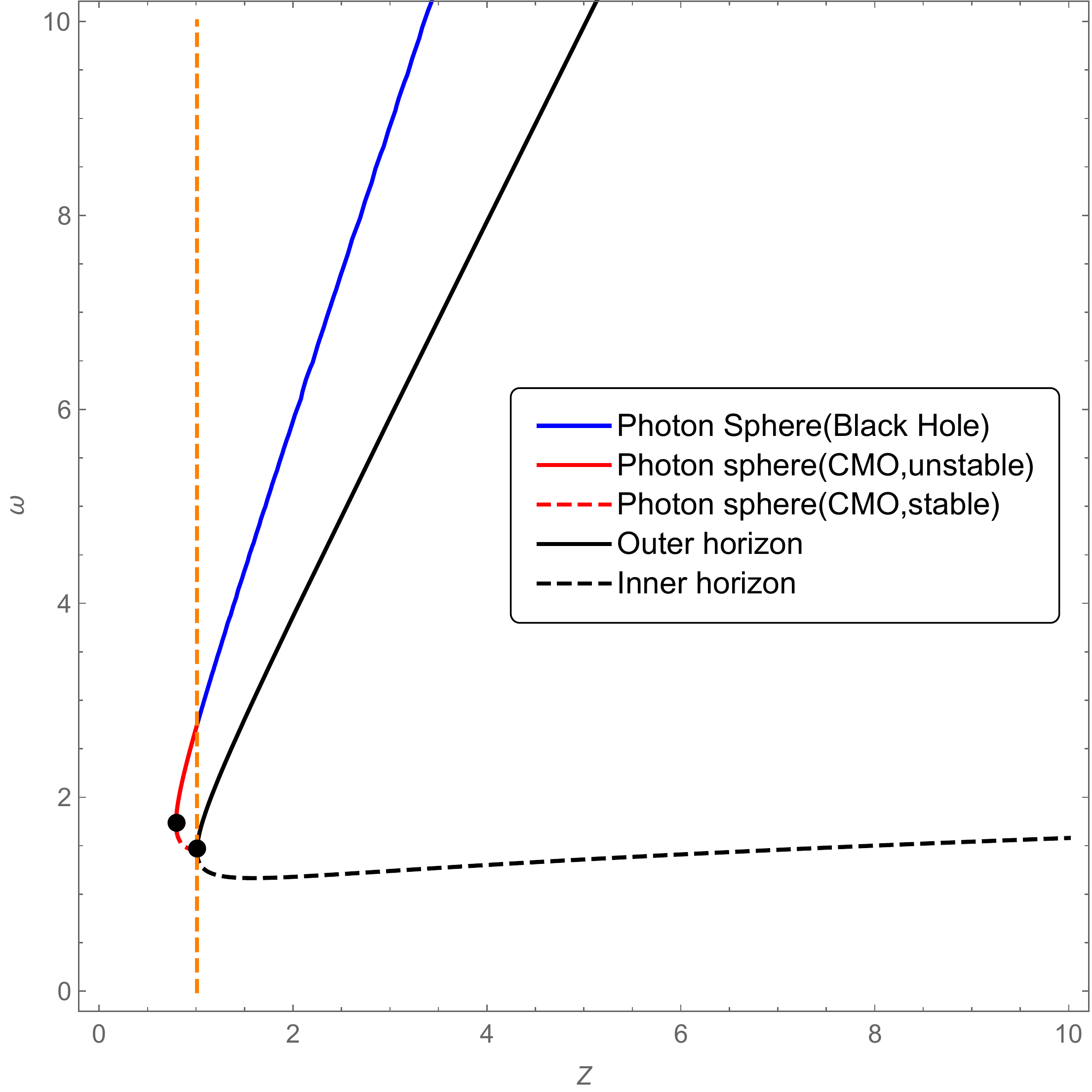}\ \hspace{0.05cm}
 \caption{\label{fig6} The location of photon spheres in the spacetime with x = 1 and n = 3. The blue line
represents the location of photon spheres in black hole phase, while the red solid line is the location of unstable photon spheres
in CMO phase, and the red dashed line is the location of stable photon spheres in CMO phase. The turning
point of each curve is marked by a black dot as well.}}
\end{figure}

The effective potential to determine  the geodesic trajectory of photons is given by
\begin{equation}
V_0(\omega)=\left(1-\frac{2z e^\frac{-z}{\omega^3}}{\omega}\right)\left(\frac{L^2}{\omega^2}\right).
\end{equation}
Setting $V_{0}'(\omega_c)=0$, one finds the radius of circular photon orbits is determined by
\begin{equation}
-3z\omega_{c}^3+e^\frac{z}{\omega_{c}^3}\omega_{c}^4+3z^2=0.
\end{equation}
Similarly, the stability of photon spheres requires $V_{0}''(\omega_c)\geq0$, leading to
\begin{equation}
-4z\omega_{c}^6+e^\frac{z}{\omega_{c}^3}\omega_{c}^7+10z^2 \omega_{c}^3-3z^3 \geq 0.
\end{equation}
We obtain the turning point by $dz/d\omega_{c}=0$, which gives rise to
\begin{equation}
z=\frac{M}{\alpha_0^{1/2}}=0.794...;\quad \omega=\frac{r}{\alpha_0^{1/2}}=1.731....
\end{equation}
We plot the radius of photon spheres as the function of the mass in Fig.(\ref{fig6}).

\subsection{Circular orbits for massive particles}
For massive particles, the effective potential is given by
\begin{equation}
V_{-1}(\omega)=\left(1-\frac{2z e^\frac{-z}{\omega^3}}{\omega}\right)\left(\frac{L^2}{\omega^2}+1\right).
\end{equation}
\begin{figure} [tbp]
 \center{
 \includegraphics[width=0.6\textwidth,height=0.6\textwidth]
{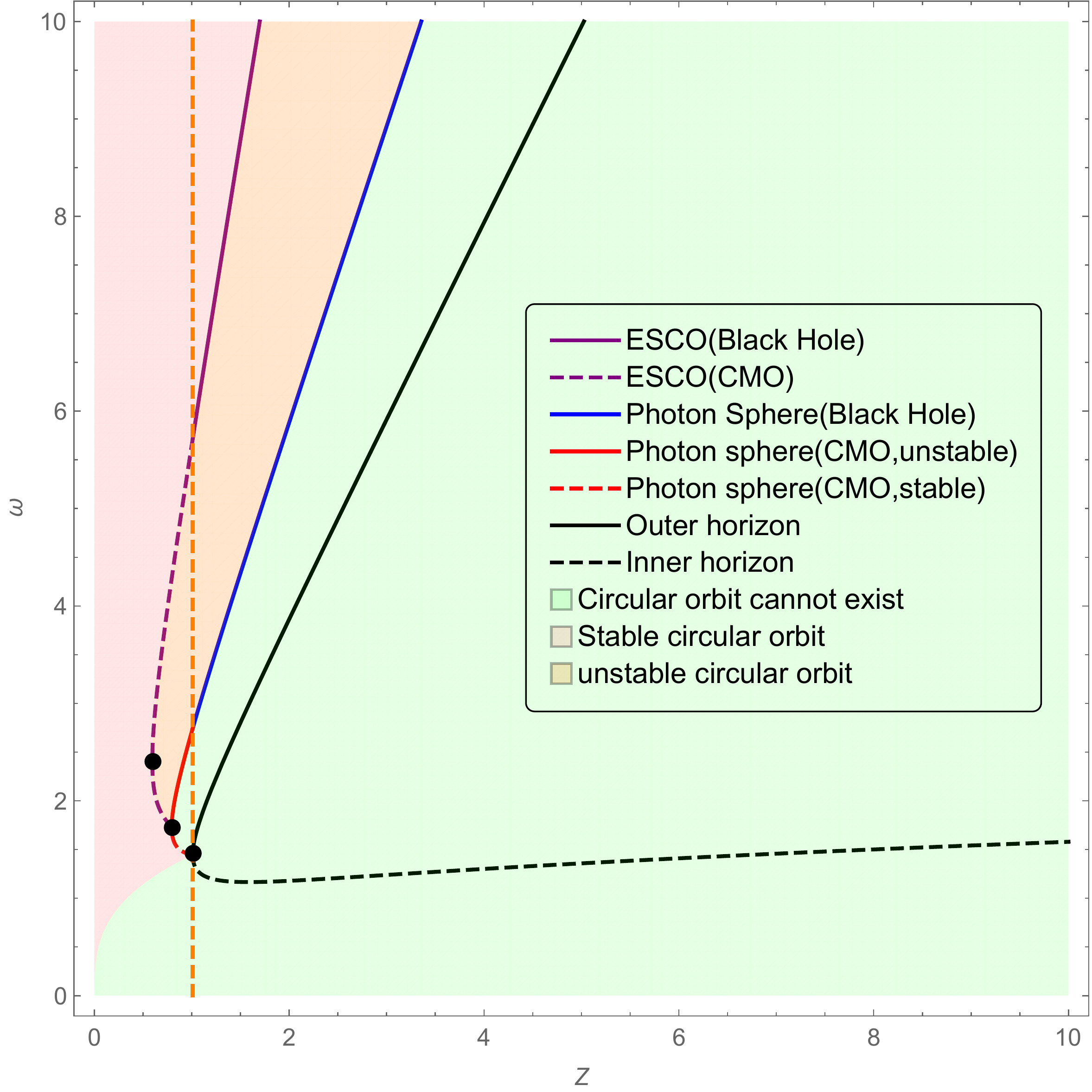} \ \hspace{0.05cm}
  \caption{\label{fig7} The location of ESCO for massive particles in the spacetime with $x=1$ and $n=3$. The purple solid line represents the location of ESCO in black hole phase, while the purple dashed line is the ESCO in CMO phase, where the upper branch is ISCO and the lower branch is OSCO.  In the region shaded in orange, all the circular orbits are unstable, while in the region shaded in pink, all the circular orbit are stable. The region shaded in green denotes the place where the circular orbit does not exist. The turning point of each curve is marked by a black dot as well.}}
\end{figure}
Then the radius of circular orbits $\omega_{e}$ is determined by $V_{-1}'(\omega_{e})=0$, which gives
\begin{equation}
z\omega_{e}^2\left(\omega_{e}^3-3z\right)-L^2\left(-3z\omega_{e}^3+e^\frac{z}{\omega_{e}^3}\omega_{e}^4+3z^2\right)=0.
\end{equation}
Furthermore, the angular momentum for all circular orbits must satisfy
\begin{equation}
L^2=\frac{z\omega_{e}^2\left(\omega_{e}^3-3z\right)}{-3z\omega_{e}^3+e^\frac{z}{\omega_{e}^3}\omega_{e}^4+3z^2}\geq 0,
\end{equation}

which gives the region where the circular orbit for massive objects is allowed. The location of ESCO is given by
\begin{equation}
e^\frac{z}{\omega_{e}^3}\omega_{e}^7-18z^3+3z^2\omega_{e}\left(2\omega_{e}^2-3e^\frac{z}{\omega_{e}^3}\right)+z\left(-6\omega_{e}^6+9e^\frac{z}{\omega_{e}^3}\omega_{e}^4\right)= 0,
\end{equation}
which gives a curve in $(z, \omega)$ plane. We plot the location of ESCO as the function of mass in Fig.(\ref{fig7}). Keep going further, the turning point of the curve of ESCO is calculated by $dz/d\omega_{e}=0$, which turns out to be
\begin{equation}
z=\frac{M}{\alpha_0^{1/2}}=0.595...;\quad \omega=\frac{r}{\alpha_0^{1/2}}=2.403....
\end{equation}

\subsection{In comparison with Hayward black hole}
The effective potential of Hayward black hole with de-Sitter core is
\begin{equation}
V_\epsilon(\omega)=\left(1-\frac{2z\omega^2}{\omega^3+ z}\right)\left(\frac{L^2}{\omega^2}-\epsilon\right).
\end{equation}
\begin{figure} [t]
  \center{
  \includegraphics[width=0.6\textwidth,height=0.6\textwidth]{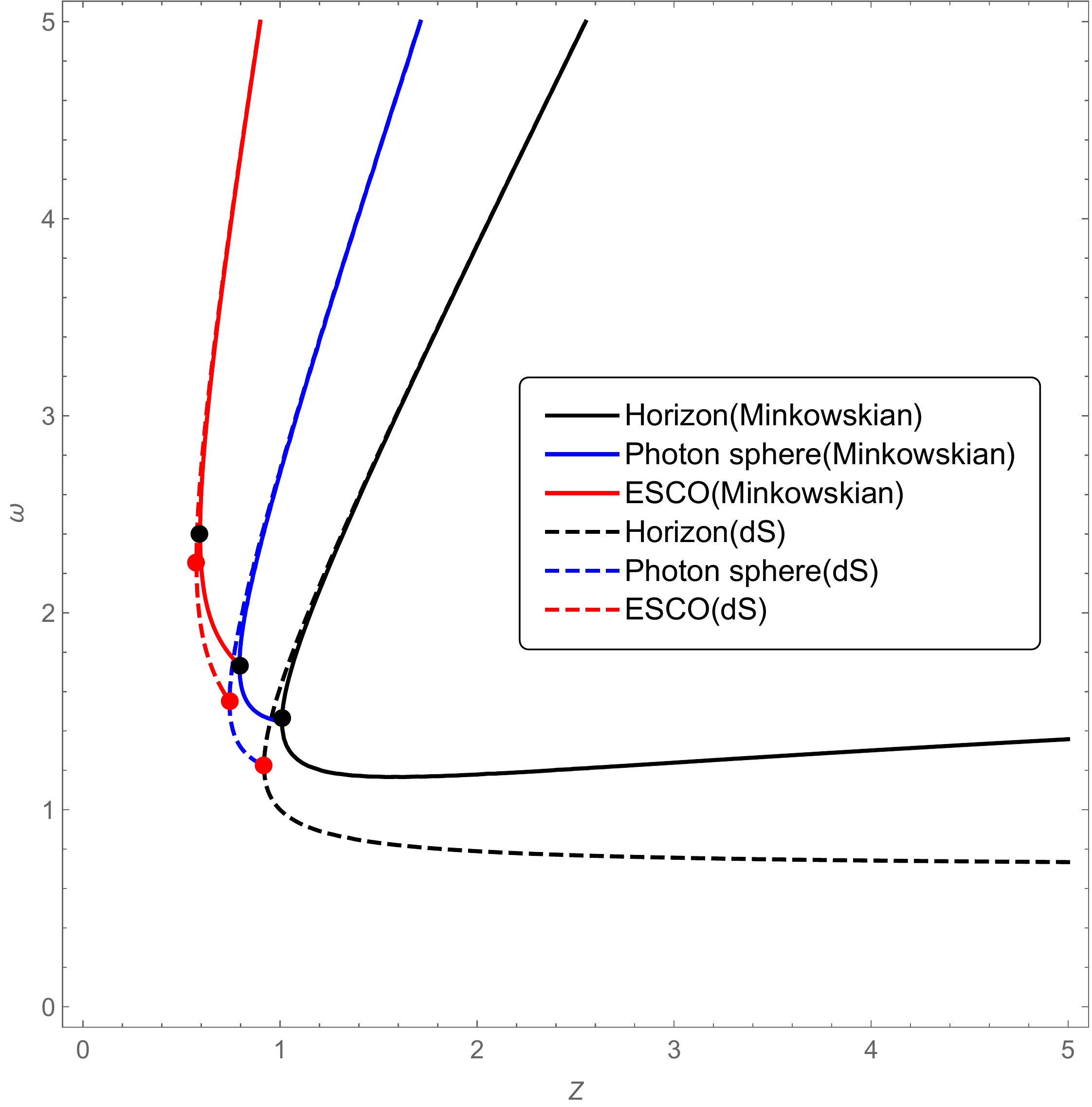}\ \hspace{0.05cm}
  \caption{\label{fig8} The comparison of photon spheres and ESCO in the spacetime with Minkowskian core $(x=1, n=3)$  and those in Hayward spacetime with dS core.} }
\end{figure}
For null trajectories, the radius of photon spheres is given by $V_{0}'(\omega_{c_{2}})=0$, leading to
\begin{equation}\label{Eq63}
\omega_{c_{2}}^5\left(-3z+\omega_{c_{2}}\right)+2z\omega_{c_{2}}^3+z^2=0.
\end{equation}
For massive particles, the ESCO is given by
\begin{equation}
    -8z^2+11z\omega_{e_2}^3+\omega_{e_2}^5\left(-6z+\omega_{e_2}\right)=0.
\end{equation}
Previously, the similar discussion on circular orbits in Hayward spacetime can be found in \cite{Chiba:2017nml}. Now we compare photon spheres as well as the ESCO in these two different spacetimes in Fig.(\ref{fig8}).
In black hole phase, the radius of circular orbits for photons and massive particles can be expanded as the series of the deviation parameter $\alpha$ as follows (The left equation is for Hayward spacetime and the right equation is for the spacetime with $(x=1,n=3)$).\\ Photon sphere:
\begin{equation}
    r_{c_2}=3M+\frac{2}{9M}\alpha_0+\mathcal{O}[\alpha_0]^2;\quad r_{c}=3M-\frac{4}{3M}\alpha_0+\mathcal{O}[\alpha_0]^2.
\end{equation}
ESCO:
\begin{equation}
    r_{e_{2}}=6M+\frac{11}{36M}\alpha_0+\mathcal{O}[\alpha_0]^2;\quad
    r_{e}=6M+\frac{19}{M}\alpha_0+\mathcal{O}[\alpha_0]^2.
\end{equation}
It is noticed that the radius of the photon sphere in Hayward black hole is larger than the standard result $3M$ in Schwarzschild black hole, while that in spacetime with $(x=1, n=3)$ is smaller than $3M$. The radius of ESCO in both spacetimes is larger than standard result $6M$ in Schwarzschild black hole. In CMO phase, both photon spheres and ESCO  exhibit distinct behavior in these two spacetimes. In particular, setting $dz/d\omega_{e}=0$ and $dz/d\omega_{c}=0$, we can get the turning point of the curves for photon sphere and ESCO in two different spacetimes, we collect these results in table (\ref{table1}).

\begin{table}[htbp]
\caption{The turning points of the curves representing the location of  horizon, photon sphere as well as ESCO in $(z,\omega)$ plane for various regular black holes.}
\label{table1}
\begin{tabular}{cccccccccc}
     \hline
      & Horizon& \quad\quad \quad\quad \quad\quad \quad\quad Photon sphere& \quad\quad\quad\quad \quad\quad\quad ESCO\\
     \hline
     $n=2, x=1$&\quad\quad ($1.359$, $1.649$)& \quad\quad\quad\quad \quad\quad \quad\quad ($1.050$, $1.953$)&\quad\quad \quad\quad \quad\quad \quad\quad ($0.806$, $2.660$)\\
     $n=2, x=2/3$&\quad\quad ($1.259$, $1.527$)& \quad\quad \quad\quad \quad\quad \quad\quad($1.037$, $1.929$)& \quad\quad \quad\quad \quad\quad \quad\quad($0.850$, $2.807$)\\
     $Bardeen$ & \quad\quad ($1.092, 1.189$)&\quad\quad \quad\quad \quad\quad \quad\quad ($0.927, 1.592$)& \quad\quad \quad\quad \quad\quad \quad\quad($0.788, 2.424$)\\
     $n=3, x=1$&\quad\quad ($1.010$, $1.465$)& \quad\quad \quad\quad \quad\quad \quad\quad($0.794$, $1.731$)&\quad\quad \quad\quad \quad\quad \quad\quad ($0.595$, $2.403$)\\
     $Hayward$ &\quad\quad($0.918, 1.224$)& \quad\quad \quad\quad \quad\quad \quad\quad($0.743, 1.549$)& \quad\quad \quad\quad \quad\quad \quad\quad($0.576, 2.258$)\\
     \hline
\end{tabular}
\end{table}
\section{CONCLUSION AND DISCUSSION}
In this paper we have investigated the circular orbits for photons and massive particles in regular black holes with sub-Planckian curvature and Minkowskian core, which exhibit abundant structure and are contrasted with the results in standard Schwarzschild black hole. It is found that in black hole phase, the location of photon sphere as well as ESCO is singe-valued, while in CMO phase it is double-valued. We have also studied the stability of these circular orbits. In black hole phase, the photon sphere is unstable, while in CMO phase the upper branch is unstable but the lower branch is stable. The ESCO has two branches in CMO phase as well. The upper one is ISCO while the lower one is OSCO. In the region between the curve of photon sphere and the curve of ESCO, all the circular orbits for massive particles are unstable, while outside the curve of ESCO, all the circular orbits are stable. We have also compared the locations of photon sphere and ESCO in spacetime with Minkowskian core and those in spacetime with dS core. In black hole phase, we have found the radius of the photon sphere in spacetime with Minkowskian core is smaller than that in spacetime with dS core. In CMO phase, we have calculated the turning points of the curves of photon sphere as well as ESCO, and found these points are distinct in these two different sorts of spacetimes. The work in this paper provides the theoretical basis for the possible detection of regular black holes in astrophysical observation.

As the first step, we have only studied the astrophysical observables in spherically symmetric spacetime. Definitely one could keep going on and extend the analysis to rotating Kerr-like black holes with sub-Planckian curvature. In this circumstance, the null geodesics with circular orbit near the horizon of the black hole has been derived and the shadow of black holes has been plotted in \cite{Ling:2022vrv}. Nevertheless, the detailed analysis on the circular orbits for massive particles and the CMO phase as the remnant of the evaporation of black holes are still absent.
In addition, to conjecture that the remnant of the evaporation could be a candidate of dark matter, the stability of horizionless CMO should be considered seriously, thus it is very worthy of studying the quasinormal modes of this sort of spacetime in CMO phase.
\section*{Acknowledgments}

We are very grateful to Meng-He Wu for helpful discussions.
This work is supported in part by the Natural Science Foundation
of China under Grant No.~11875053 and 12035016. It is also supported by Beijing Natural Science Foundation under Grant No. 1222031, and by Sichuan Youth Science and Technology Innovation Research Team with Grant No.~21CXTD0038, by Central Government Funds of Guiding Local Scientific and Technological Development for Sichuan Province with Grant No. 2021ZYD0032.

\end{document}